\documentclass[runningheads]{llncs}
\usepackage[T1]{fontenc}
\usepackage{graphicx}
\usepackage{floatrow}
\newfloatcommand{capbtabbox}{table}[][\FBwidth]

\usepackage[misc,geometry]{ifsym}
\usepackage{float}
\usepackage{hyperref}
\usepackage{amsmath}
\usepackage{multicol,multirow}
\usepackage{booktabs}
\usepackage{color}

\begin{document}
\title{Graph Convolutional Neural Networks to Model the Brain for Insomnia}
%
%
\author{Kevin Monteiro\inst{1}\textsuperscript{(\Letter)}
\and 
Sam Nallaperuma-Herzberg\inst{1} \and 
Martina Mason\inst{2} \and 
Steve Niederer\inst{3}}
\authorrunning{K. Monteiro et al.}
%
\institute{University of Cambridge, Cambridge, UK \\
\email{kidrm2@cam.ac.uk} \and
Respiratory and Sleep Centre, Royal Papworth Hospital, Cambridge, UK \and
Alan Turing Institute, London, UK}
\maketitle              
\begin{abstract}
Insomnia affects a vast population of the world and can have a wide range of causes. Existing treatments for insomnia have been linked with many side effects like headaches, dizziness, etc. As such, there is a clear need for improved insomnia treatment. Brain modelling has helped with assessing the effects of brain pathology on brain network dynamics and with supporting clinical decisions in the treatment of Alzheimer’s disease, epilepsy, etc. However, such models have not been developed for insomnia. Therefore, this project attempts to understand the characteristics of the brain of individuals experiencing insomnia using continuous long-duration EEG data. Brain networks are derived based on functional connectivity and spatial distance between EEG channels. The power spectral density of the channels is then computed for the major brain wave frequency bands. A graph convolutional neural network (GCNN) model is then trained to capture the functional characteristics associated with insomnia and configured for the classification task to judge performance. Results indicated a 50-second non-overlapping sliding window was the most suitable choice for EEG segmentation. This approach achieved a  classification accuracy of 70\% at window level and 68\% at subject level. Additionally, the omission of EEG channels C4-P4, F4-C4 and C4-A1 caused higher degradation in model performance than the removal of other channels. These channel electrodes are positioned near brain regions known to exhibit atypical levels of functional connectivity in individuals with insomnia, which can explain such results.

\end{abstract}
\section{Introduction}
Insomnia is a common sleeping disorder that affects around a third of the adult population world-wide \cite{insomnia-update,aernout2021international}.
Many cases of insomnia also go unreported and it is estimated that the fraction of individuals who seek medical advice is as low as 30\% (\href{https://cks.nice.org.uk/topics/insomnia/background-information/prevalence/}{National Institute for Health and Care Excellence}). 
The underlying causes for insomnia vary widely from case to case including mental health disorders such as anxiety and depression, physical illnesses such as migraine or cancer, medications, hormonal changes occurring in the menstrual cycle, pregnancy and menopause, neurological problems, etc. (\href{https://aasm.org/resources/factsheets/insomnia.pdf}{The American Academy of Sleep Medicine}). 

There has been a growing interest in brain modelling research to construct predictive models for brain function and dysfunction enabling cheaper hypotheses testing in silico. 
As such, this work aims to develop a model to capture the characteristics of a human brain that is affected by insomnia and which is enriched with the foundations of neuroscience, cognitive psychology and network theory as well as real-world data. Preliminary models have been proposed in the state-of-the-art for example, for Alzheimer’s disease \cite{digi-twin-alzheimers-bertolini2020modeling} and multiple sclerosis \cite{digi-twin-multiple-sclerosis}. However, there are no comprehensive brain models available yet to capture the psychological aspects which affect insomnia.

Background noise and the non-stationary random nature of the EEG signal makes analysis difficult however methods have advanced over recent years. Graph neural networks (GNN) \cite{klepl2022eeg} in particular, have shown great promise 
while handling non-euclidean graph-structured EEG data  
where graphical representations for the EEG segments were computed based on measures (such as correlation between the channels) which corresponded to edge weight/importance. These graphs were then passed to a GNN model as edge features while the power spectral density (PSD) of the channels were used as node features to carry out the classification task for Alzheimer’s disease.
A similar approach was followed by \cite{wagh2020eeg} for the detection of epilepsy through EEG analysis.

Studies suggest that an increase in Beta frequency band activity was related to higher arousal levels which could manifest as insomnia \cite{zhang2022quantitative}.
Furthermore, works like \cite{killgore2013insomnia,chen2014increased,wang2017increased} uncovered that insomnia was associated with elevated levels of functional connectivity between specific regions of the brain when compared to control subjects, thus, providing evidence of learnable features to model the brain of an insomnia patient.

This work makes the following main contributions.
  1) Presentation of an overview of measures that can be used to estimate functional connectivity and construct brain networks.
%
  2) Development of a novel graph convolutional neural network 
  model that captures the characteristics of a human brain experiencing insomnia for which no such model has been put forth yet.
%
%
  3) Identifying effective features for the detection of insomnia characteristics in EEG recordings which are derived from functional connectivity and spatial distance.
%
  4) Exploration of the importance of each EEG channel in the apprehension of insomnia characteristics for the insomnia detection task.
  

\section{Background}
\subsection{Functional Connectivity}
Functional connectivity (FC) is a concept that can be used to construct brain graphs/networks. It is defined as the statistical interdependence between spatially separated neuro-physiological regions of the brain.
Over the past few years, brain imaging techniques like functional Magnetic Resonance Imaging (fMRI), magnetoencephalography (MEG) and electroencephalography have been used to derive functional connectivity measures to study the brain. 

Researchers have calculated cortical connectivity through parametric approaches 
like the Directed Transfer Function (DTF),
Partial Directed Coherence (PDC)
and the direct DTF (dDTF). 
Comparing these estimates at different signal-to-noise ratios showed that the error in connectivity estimates was less when using DTF based methods as compared to PDC which suggests that DTF based methods provide better results when working with noisy signals. However, as the DTF was unable to differentiate between indirect and direct causal flows, significant values were attributed to indirect paths. This leads to bias in the FC estimates. While dDTF solves this problem to a certain extent it does still exhibit higher bias than PDC and outputs lower connectivity values overall as compared to DTF and PDC \cite{astolfi2007comparison}.

Among the nonparametric methods, measures based on cross-correlation between signals produced by spatially distributed brain regions have been leveraged in the time domain for cortical connectivity estimation \cite{lee2014classifying}. 
EEG coherence, the counterpart of correlation in the frequency domain 
is also widely used in literature \cite{srinivasan2007eeg}. 
To reduce the effect of volume conduction when using EEG coherence, \cite{nunez1997eeg} proposed approximating the true cortical region coherence (Reduced Coherence) by subtracting random coherence from the measured coherence.
Another means to partially mitigate EEG coherence caused by volume conduction is to reduce the spatial scale of electrode potentials with high-resolution EEG techniques (e.g. Surface Laplacian) \cite{nunez1997eeg}. 
However, in the case of activity sources with low spatial frequencies (i.e. source regions that are widely distributed), Laplacians can remove true source contributions along with volume conduction effects.
Time-frequency analysis based techniques have also been used to estimate functional connectivity.
Wavelet coherence (WC) enables the simultaneous spectral and temporal analysis of EEG channel synchronisation \cite{sankari2012wavelet}. Short-time Fourier coherence (STFC) \cite{wendling2009eeg} uses a sliding window of fixed size and conducts spectral analysis over the duration covered by the window however a fixed window size is not always optimal for various EEG recording frequencies.
Other nonparametric techniques include non-linear connectivity measures like 
the phase locking value (PLV) \cite{bajo2015scopolamine} and the phase lag index (PLI) \cite{chaturvedi2019phase} from which phase synchronisation strengths can be derived. Information theory also provides techniques from which non-linear connectivity values can be obtained. These are mutual information (MI) \cite{melia2015mutual}, synchronisation likelihood (SL) \cite{chriskos2018achieving} and transfer entropy (TE) \cite{harmah2020measuring}. MI and SL provide undirected functional connectivity estimates while TE has the ability to indicate the direction of information flow.

Parametric techniques usually have a compact and transparent model structure. Their variables are generally lagged EEG signals and they do not require a large number of samples. On the other hand, nonparametric techniques are generally model-free and require large quantities of data to provide good estimates of cortical connectivity.

\subsection{Existing Brain Networks in Sleep Studies}
Advancements in technology and research have facilitated brain network development. These networks have been adopted in various sleep studies with great success, although, most studies have focused on constructing brain networks using functional Magnetic Resonance Imaging (fMRI) scans.

\cite{huang2021eeg} tackled the sleep staging problem using the phase locking value (PLV) to estimate the connectivity matrix of the various brain regions. This was done for six different brain frequency bands extracted from the EEG recording, each of which had a corresponding brain graph. The entries in the connectivity matrix, which approximated the synchronisation between the considered brain regions were passed through a threshold that was set as the highest value at which the brain network did not have any isolated nodes. Fusion strategies were employed to combine the feature information from the six frequency bands before performing classification with a support vector machine (SVM). 

Mutual information (MI) is another attractive  brain connectivity measure in this regard as demonstrated in \cite{aydin2015mutual}. Psycho-physiological insomnia patients exhibited brain networks with lower MI values as compared to the control group. Classification performance was then used to judge the performance of the MI measure.

\cite{jia-2020graphsleepnet} proposed a novel GNN approach that utilised a spatial-temporal graph convolution network (ST-GCN) for sleep stage classification. In this architecture, EEG channels were represented by graph nodes and connections between channels which were adaptively learnt corresponded to graph edges. The ST-GCN used graph convolutions to obtain spatial features and temporal convolutions to capture sleep stage transition features. 
Then, the attention mechanism was incorporated to provide weights to the features by importance. 
Using this approach state-of-the-art performance was achieved in the sleep stage classification task.

Based on research in \cite{killgore2013insomnia}, \cite{chen2014increased} and \cite{wang2017increased} illustrating how the connectivity pattern of the brain regions in individuals experiencing insomnia can differ from normal subjects,  \cite{zhang2022quantitative} introduced an approach to detect insomnia using brain networks of patients in resting state. To increase the robustness of the brain network, three functional connectivity estimated were considered which were Pearson Correlation Coefficient (PCC), phase lag index (PLI) and Partial Directed Coherence (PDC). To decrease network complexity and remove insignificant edges, the connectivity matrix 
was binarized based on a threshold. 
%
The largest possible threshold value was computed to decrease node isolation and randomness in graph topology. To evaluate the derived brain network, the clustering coefficient $C$, the average shortest path length between all node pairs $L$, and global efficiency $E_\textbf{global}$ defined as the sum of communication efficiencies of all nodes 
were combined to form an index $\eta$. 
This index was computed for each sliding window. A high $\eta$ value indicated more ``small-world" properties in the brain network. The index value computed was found to be higher in the group of insomnia patients as compared to the normal control group in the case of all three connectivity estimates. The brain network features were then fed into a bidirectional LSTM model for classification.

\section{Proposed Approach}

\subsection{Dataset}
The CAP (Cyclic Alternating Pattern) Sleep Database \cite{terzano2001atlas}, available on PhysioNet \cite{goldberger2000physiobank}, was considered in this project. 
Of all the polysomnographic recordings, 9 recordings were of insomnia patients while 16 recordings were of healthy control individuals. 
Due to inconsistency in the recorded channels and data corruption in some recordings, only 7 recordings from the control group could be used in the experiments. 
Each recording was approximately 13 hours long, sampled at 512 Hz. The final dataset contained recordings from 6 male and 10 female subjects. The average age of the subjects was 48.53 with a standard deviation of 17.25.

\subsection{EEG Signals Pre-processing} \label{EEG Signals pre-processing}

As there are no noteworthy brain waves lower than 1 Hz
, the EEG signals were passed through a high-pass filter (with a cutoff of 1 Hz). The signals were further denoised using a notch filter and down-sampled to 250 Hz to facilitate quicker processing.
In line with most sleep-related EEG pre-processing routines, a uniform segmentation scheme in which
the EEG time course was broken down into non-overlapping segments using a sliding window approach was followed. There is always a trade-off when selecting a window length due to the non-stationary nature of the EEG signal. However, it is generally assumed that the signal remains quasi-stationary over short durations of time.
Window durations vary in literature and can range from a couple of seconds to around 60 seconds in sleep studies.
To judge the effect of the window length on the model, multiple experiments were conducted with the window length set as 10, 30, 50, 70 and 90 seconds.

\subsection{Brain Network Construction and Feature Extraction} \label{Brain network construction and feature extraction}
We model the brain network as a graph $G = (V, E)$ where the nodes in $V$ corresponded to the EEG channels (sensor-signal approach), which in turn acted as proxies for brain regions. 
The graph node positions were set as the position 
in the middle of the two electrodes corresponding to the given channel.
By this mapping, the spatial separation $D_{ij}$ between the nodes/channels $i$ and $j$ was calculated as the geodesic distance between the corresponding nodes.
%
The edges in $E$ were computed based on a functional connectivity measure. More specifically, spectral coherence was used to estimate functional connectivity between the nodes. 
To account for the effect of volume conduction, the random coherence $C_{ij}^{(random)}$ \cite{nunez1997eeg} was subtracted from the computed coherence values $C_{ij}^{(computed)}$. That is, for two EEG channels $i$ and $j$, the random coherence $C_{ij}^{(random)}$ was approximated as:
\begin{equation}
    C_{ij}^{(random)} = \textbf{exp}\left(\frac{1-D_{ij}}{k}\right)
\end{equation}
where 
$k$ is a constant \cite{nunez1997eeg}.
The computed coherence $C_{ij}^{(computed)}$ was calculated based on the coherence function 
defined in \cite{bendat2011random}. 
So the (reduced) spectral coherence $C_{ij}$ \cite{nunez1997eeg} between channel $i$ and channel $j$ was obtained as:
\begin{equation}
    C_{ij} = C_{ij}^{(computed)} - C_{ij}^{(random)}
\end{equation}

To further improve the connectivity estimate, the linear contribution of the spatial distances between the electrodes was also considered based on results in \cite{wagh2020eeg}. 

\begin{figure}[H]
\includegraphics[scale=0.65]{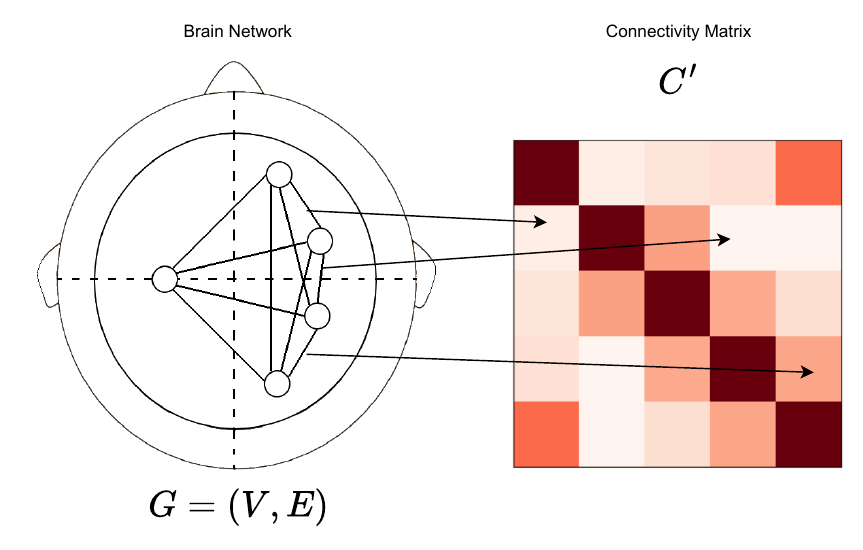}
\centering
\caption{Visual representation of the overall connectivity matrix obtained from the brain network}
\label{fig:brain-net-conn-mat}
\end{figure}

So the overall connectivity estimate $C_{ij}'$ was computed as:
\begin{equation}
C_{ij}' = C_{ij} + D_{ij} = C_{ij}^{(computed)} - C_{ij}^{(random)} + D_{ij}
\end{equation}
and then normalised to belong to the range $[0, 1]$.
As such, a 5 $\times$ 5 connectivity matrix was obtained for each window depicting cortical connectivity between the nodes (Figure \ref{fig:brain-net-conn-mat}). 

Power spectral density (PSD) was then computed using Welch’s method \cite{psd-welch} for six brain wave frequency bands (i.e. Delta (1 to 4 Hz), Theta (4 to 8 Hz), Alpha (8 to 13 Hz), lower Beta (13 to 16 Hz), higher Beta (16 to 30 Hz) and Gamma (30 to 150 Hz)) in each channel. This matrix was used as node features.

\subsection{Graph Convolutional Neural Network Model} \label{Graph convolutional neural network model}

\begin{figure}[H]
\includegraphics[scale=0.65]{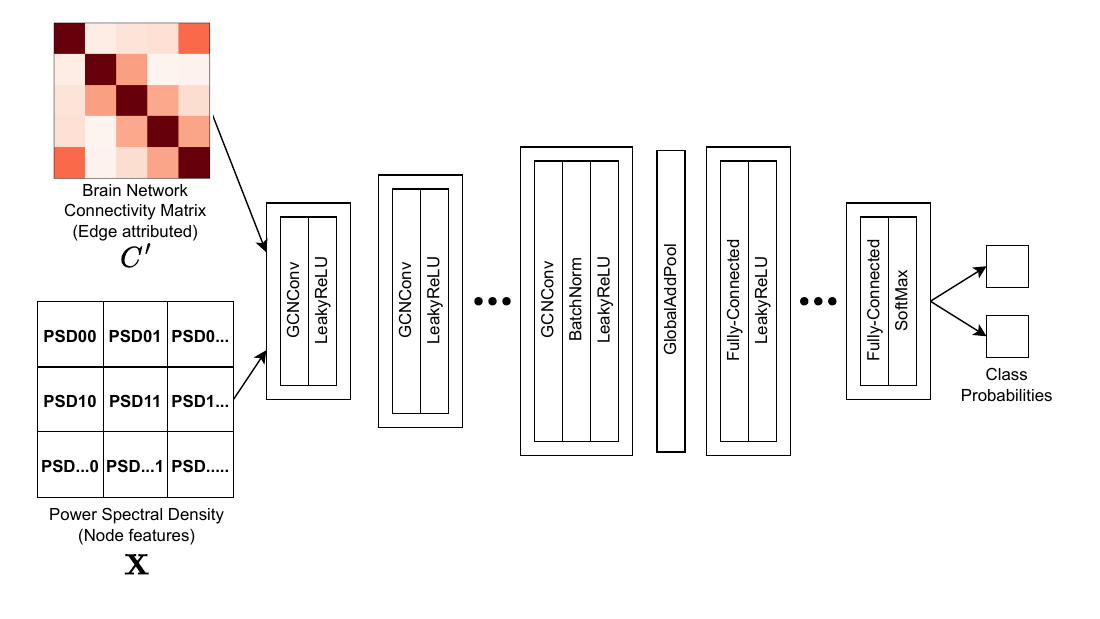}
\centering
\caption{Overview of the model architecture. The brain network as represented by its connectivity matrix and PSD matrix are supplied as input. The output is the class probabilities}
\label{fig:gcnn-arch}
\end{figure}

The graph neural network model was built using the implementation of the graph convolutional operator in \cite{kipf2016semi}. 
%
The model contained graph convolutional layers with a LeakyReLU activation. Figure \ref{fig:gcnn-arch} illustrates the overall architecture.
The connectivity matrix of the brain network was passed in as edge attributes and the PSD matrix was passed in as node features.

This graph convolutional neural network was configured for the insomnia detection task to evaluate the performance of the entire machine learning pipeline (as done in \cite{aydin2015mutual}). 
In particular, after applying the graph convolutional operation, output node features are aggregated to obtain graph level features which are subsequently passed through dense layers to obtain the class probabilities.
The accuracy, precision, recall and F1 score were then captured. Window level results were also aggregated to produce patient level prediction by computing the prediction accuracy for each patient and averaging across all.


\section{Experiments}
To quantify the performance of the models tested, a 5-fold subject-independent cross-validation strategy was followed. As such, the data was partitioned into disjoint training and testing sets based on the patient identification number, ensuring all window information of a subject belonged to only one set at a time. This was then averaged over 3 iterations to improve the reliability of the results. We train using stochastic gradient descent with an initial learning rate of 0.01 and a decay factor of 10 after every 10 epochs. 

\subsection{Results}
\subsubsection{Effect of Window Length}
\begin{figure}[H]
\includegraphics[scale=0.45]{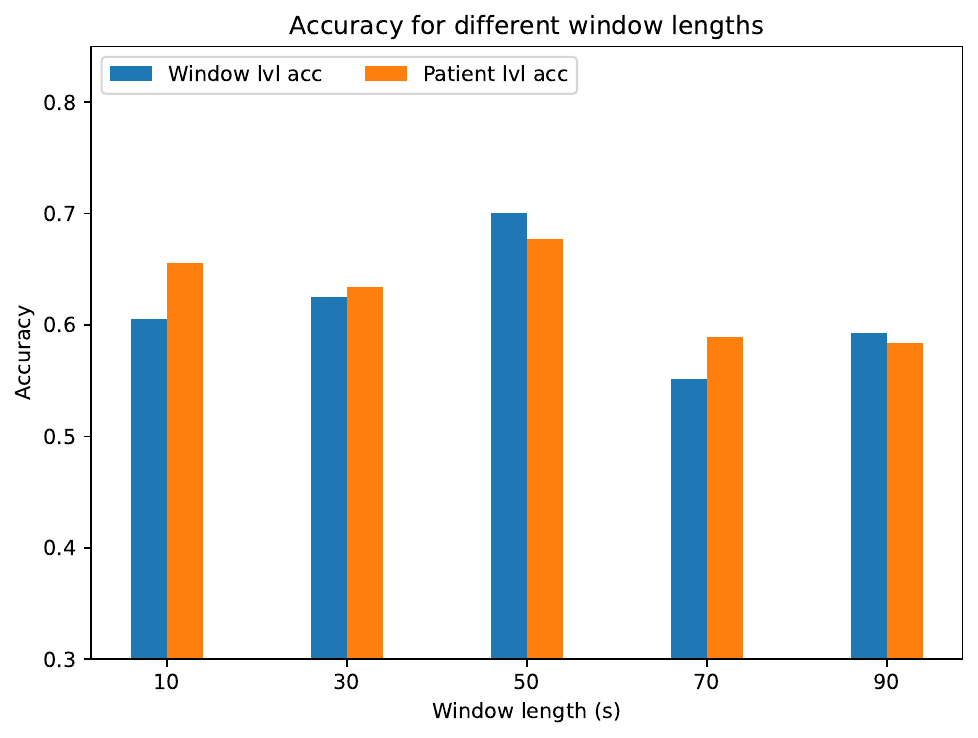}
\centering
\caption{Average window level accuracy and subject level accuracy for different window lengths}
\label{fig:acc-for-different-win-sizes}
\end{figure}

To evaluate the effect of changing the window length, multiple experiments with different window sizes were conducted. The results of these experiments are presented in Table \ref{tab:results-window-length} and Figure \ref{fig:acc-for-different-win-sizes} shows a bar-plot of the accuracy corresponding to the various window lengths.
Overall, the classification accuracy increased as the window size was made larger, reaching a maximum window level accuracy of 70.1\% and subject prediction level accuracy of 67.7\% using a window size of 50 seconds. On increasing the window size still further a decrease in accuracy was observed. 
Also, models constructed using window lengths of 10 and 30 seconds performed better than those constructed using window lengths of 70 and 90 seconds.
A CAP cycle (constituted by a phase of brain activation that is followed by deactivation) during NREM sleep is usually less than a minute \cite{terzano2001atlas}. As such, segment lengths of 10, 30 and 50 seconds are more closely able to overlap with these CAP cycles leading to fewer windows which contain partial data of two neighbouring CAP cycles. This can explain the relatively poorer performance of the models corresponding to window sizes 70 and 90 seconds.


\begin{table}[h!]
\centering
\resizebox{7.5cm}{!}{
\begin{tabular}{@{}cccccc@{}}
\toprule
\multirow{2}{*}{\textbf{\begin{tabular}[c]{@{}c@{}}Window\\length (s) \end{tabular}}}  &  \multicolumn{4}{c}{\textbf{Window level}} & \multirow{2}{*}{\textbf{\begin{tabular}[c]{@{}c@{}}Subject\\accuracy \end{tabular}}} \\ \cmidrule(l){2-5}  & \textbf{Accuracy} & \textbf{Precision} & \textbf{Recall} & \textbf{F1}  &  \\ 
 \midrule
10 & 0.605 & 0.591 & 0.604 & 0.603 & 0.656\\
\midrule
30 & 0.625 & 0.613 & 0.620 & 0.616 & 0.634\\
\midrule
\textbf{50} & \textbf{0.701} & \textbf{0.667} & \textbf{0.699} & \textbf{0.706} &  \textbf{0.677}\\
\midrule
70 & 0.552 & 0.560 & 0.551 & 0.548 & 0.589\\
\midrule
90 & 0.593 & 0.539 & 0.592  & 0.632 & 0.584\\
\bottomrule \\
\end{tabular}
}%
\caption{Average window level accuracy, precision, recall, F1 score and subject level accuracy for different window lengths. The best model is indicated in bold.}
\label{tab:results-window-length}
\end{table}

\subsubsection{Connectivity Estimate Comparison}
\begin{figure}[H]
\includegraphics[scale=0.45]{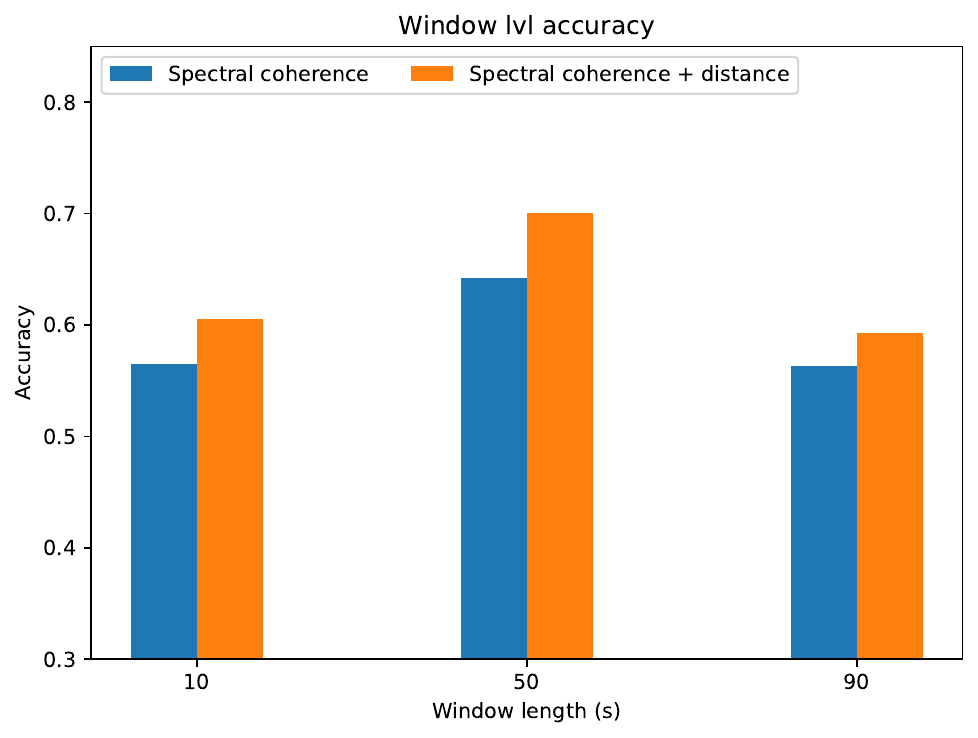}
\centering
\caption{Average window level accuracy and subject level accuracy for different connectivity measures}
\label{fig:conn-estimate-comparison}
\end{figure}

To quantify the improvement in the connectivity estimate achieved by combining the functional connectivity estimate (spectral coherence) and the spatial distance measure, two sets of tests, one including and the other excluding the spatial measure, were performed when generating the brain graph. The results of these experiments are documented in Table \ref{tab:results-connectivity}. The accuracy corresponding to the connectivity estimates in these two sets of tests is depicted in the bar plot in Figure \ref{fig:conn-estimate-comparison}. The models obtained using the combined connectivity measure outperformed models obtained considering only spectral coherence.
More specifically, 
with a window size of 50 seconds, the window level accuracy increased from approximately 64\% to 70\% and the subject level accuracy rose from 62\% to 68\% when using the combined connectivity measure.
These results suggest that while functional connectivity 
is a powerful tool to 
derive brain networks, these networks can be enhanced using 
spatial characteristics to obtain a more holistic model of brain dynamics during sleep.

\begin{table}[h!]
\centering
\resizebox{10.5cm}{!}{
\begin{tabular}{@{}ccccccc@{}}
\toprule
\multirow{2}{*}{\textbf{\begin{tabular}[c]{@{}c@{}}Window\\length (s) \end{tabular}}} & \multirow{2}{*}{\textbf{\begin{tabular}[c]{@{}c@{}}Connectivity\end{tabular}}} &  \multicolumn{4}{c}{\textbf{Window level}} & \multirow{2}{*}{\textbf{\begin{tabular}[c]{@{}c@{}}Subject\\accuracy \end{tabular}}} \\ \cmidrule(l){3-6}  & & \textbf{Accuracy} & \textbf{Precision} & \textbf{Recall} & \textbf{F1}  &  \\ 
 \midrule
\multirow{2}{*}{10}& Spectral coherence & 0.565 & 0.556 & 0.564 & 0.566 & 0.617\\
\cmidrule(l){2-7}
                   & \begin{tabular}[c]{@{}c@{}}Spectral coherence and distance \end{tabular}  & 0.605 & 0.591 & 0.604 & 0.603 & 0.656 \\
\midrule
\multirow{2}{*}{\textbf{50}}& Spectral coherence & 0.642 & 0.567 & 0.640 & 0.657 & 0.621 \\
\cmidrule(l){2-7}
                   & \begin{tabular}[c]{@{}c@{}}Spectral coherence and distance \end{tabular} & \textbf{0.701} & \textbf{0.667} & \textbf{0.699} & \textbf{0.706} &  \textbf{0.677}\\
\midrule
\multirow{2}{*}{90}& Spectral coherence & 0.563 & 0.557 & 0.561 & 0.606 & 0.567 \\
\cmidrule(l){2-7}
                   & \begin{tabular}[c]{@{}c@{}}Spectral coherence and distance \end{tabular} & 0.593 & 0.539 & 0.592  & 0.632 & 0.584\\
\bottomrule \\
\end{tabular}
}%
\caption{Average window level accuracy, precision, recall, F1 score and subject level accuracy for different connectivity measures.}
\label{tab:results-connectivity}
\end{table}

\subsubsection{EEG Channel Importance}
\begin{figure}[H]
\includegraphics[scale=0.45]{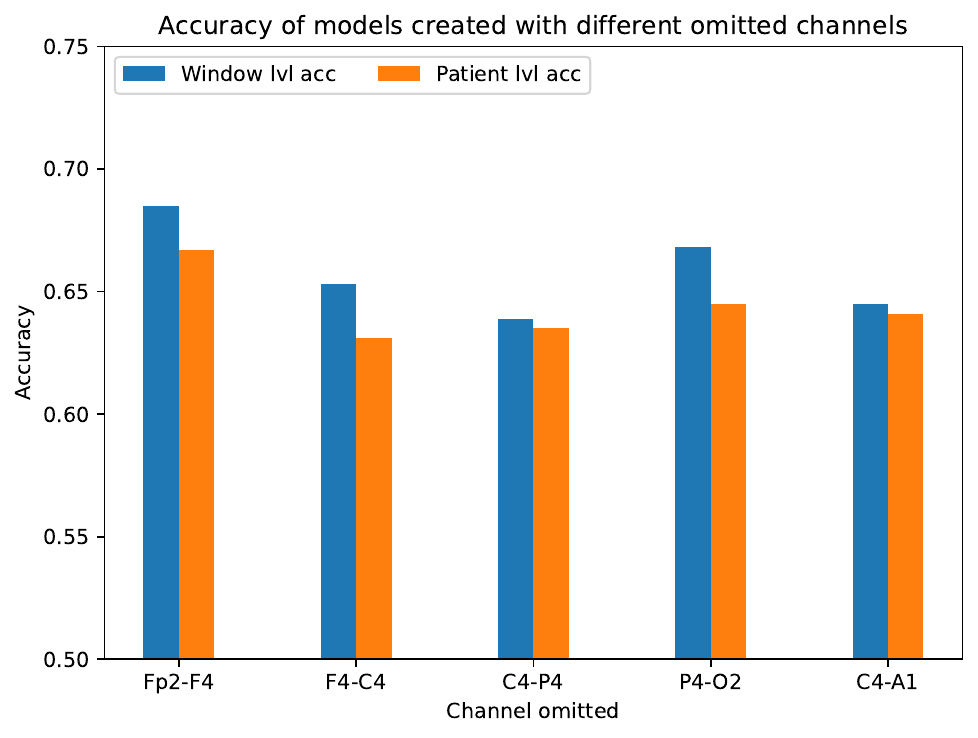}
\centering
\caption{Average window level accuracy and subject level accuracy of models in the channel importance experiments}
\label{fig:ch-imp}
\end{figure}

Understanding which brain regions are most important (while modelling the brain of an individual with insomnia) is useful to pave the way for model compression and simplification. To determine this aspect, the following experimentation protocol was followed. First, separate brain networks were constructed leaving one unique EEG channel out in each case. The resultant brain graphs contained 4 nodes each with different missing nodes (corresponding to the omitted channel). Graph convolutional neural network models were then trained for each case and the channel importance was assessed based on the amount of performance loss in the classification task.
Table \ref{tab:results-single-channel-removed} provides a quantitative overview of the performance of each model while Figure \ref{fig:ch-imp} presents a bar plot of the accuracy of each model.

\begin{table}[h!]
\centering
\resizebox{7cm}{!}{%
\begin{tabular}{@{}cccccc@{}}
\toprule
\multirow{2}{*}{\textbf{\begin{tabular}[c]{@{}c@{}}Channel\\omitted \end{tabular}}}  &  \multicolumn{4}{c}{\textbf{Window level}} & \multirow{2}{*}{\textbf{\begin{tabular}[c]{@{}c@{}}Subject\\accuracy \end{tabular}}} \\ \cmidrule(l){2-5}  & \textbf{Accuracy} & \textbf{Precision} & \textbf{Recall} & \textbf{F1}  &  \\ 
 \midrule
Fp2-F4 & 0.685 & 0.565 & 0.686 & 0.677 & 0.667\\
\midrule
F4-C4 & 0.653 & 0.579 & 0.655 & 0.668 & 0.631\\
\midrule
C4-P4 & 0.639 & 0.564 & 0.638 & 0.654 & 0.635\\
\midrule
P4-O2 & 0.668 & 0.544 & 0.667 & 0.662 & 0.645\\
\midrule
C4-A1 & 0.645 & 0.643 & 0.646 & 0.669 & 0.641\\
\bottomrule \\
\end{tabular}
}%
\caption{Average window level accuracy, precision, recall, F1 score and subject level accuracy of models where the channel omitted during brain network creation is indicated in the leftmost column.}
\label{tab:results-single-channel-removed}
\end{table}

\begin{figure}[H]
\includegraphics[scale=0.27]{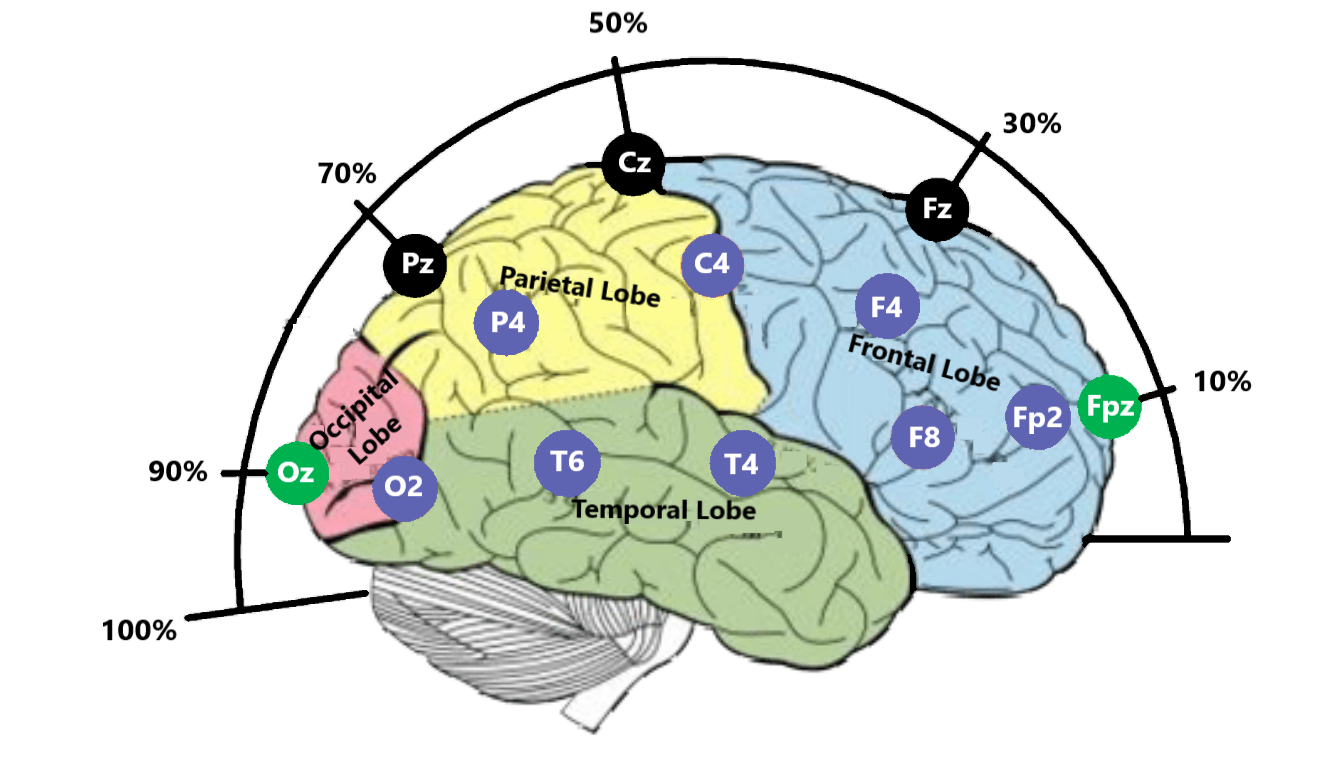}
\centering
\caption{EEG electrode placement according to the 10-20 system over the lobes of the right hemisphere of the brain}
\label{fig:brain-right}
\end{figure}

As can be seen, the removal of channel C4-P4 caused the highest adverse impact to model performance (i.e. a drop in window level accuracy from 70.1\% to 63.9\% and a drop in subject level accuracy from 67.7\% to 63.5\%)
while the removal of channel Fp2-F4 caused the least loss in performance (i.e. a drop in window level accuracy from 70.1\% to 68.5\% and a drop in subject level accuracy from 67.7\% to 66.7\%). Thus, the three most important channels were found to be C4-P4, F4-C4 and C4-A1. 
These results were  consistent with the work in  \cite{killgore2013insomnia},  \cite{chen2014increased} and  \cite{wang2017increased} which demonstrated that parts of the motor cortex, auditory cortex, insular and sensory areas exhibited abnormal levels of functional connectivity in brains affected by insomnia.
The position of the EEG electrodes with respect to the lobes that are made up by the aforementioned regions is illustrated in Figure \ref{fig:brain-right}.
Electrode C4 is placed above the motor cortex (located at the posterior end of the frontal lobe) and is near the insular cortex (located deep within the centre of the brain, in the fissure separating the temporal lobe from the parietal and frontal lobes). Electrode P4 is placed above the parietal lobe which is concerned with sensory perception and integration, including the management of taste, hearing, sight, touch, and smell. Electrode A1 is placed above the left ear, near the auditory cortex (which is part of the temporal lobe)
As such it is reasonable to conclude that channels derived from electrodes placed very close to regions of the brain that demonstrate deviation from normal functional connectivity (in subjects affected by insomnia) would have a greater influence on modelling and differentiating ability.

\section{Conclusion}

In the domain of insomnia research, there is a very limited body of work that constructs brain graphs/networks from EEG data. Moreover, in studies that consider EEG as the sole modality, the recordings used are usually of brain activity during short periods of interest such as before sleep onset and after waking.
This study 
uses continuous long-duration EEG recordings (of about 13 hours each) to learn the evolution in brain dynamics during sleep.
We employ spectral coherence, augmented with spatial distance features to estimate brain connectivity.
A graph convolutional neural network model was then constructed based on the implementation in \cite{kipf2016semi}
and was configured for the classification task to assess performance.

With a window size of 50 seconds, the best classification performance was observed which corresponded to a window level prediction accuracy of around 70\% and a patient level prediction accuracy of approximately 68\%. 
Additionally, models trained with window lengths that could correspond to possible CAP cycle durations (10, 30 and 50 seconds) performed better in general.
Channel importance experiments 
revealed that channel C4-P4 was the most important as excluding it resulted in the largest drop in the model performance.

These findings were in accordance with the clinical studies which uncovered that brain regions located very close to the electrode that constituted the more important channels showed atypical levels of functional connectivity in individuals suffering from insomnia.

In conclusion, this work provides an overview and valuable insights into the application of functional connectivity to develop brain networks for insomnia.  Future work will focus on extending this approach to a multi-modal model, incorporating other brain data modalities such as fMRI and MEG.

\begin{credits}
\subsubsection{\ackname} 
The authors thank Prof. Carl Henrik Ek and Prof. Pietro Lio for their valuable insights. This work is jointly funded by the Accelerate Program for Scientific Discovery and BrainTwin project grant, University of Cambridge.

\subsubsection{\discintname}
The authors have no competing interests to declare that are relevant to the content of this article.
\end{credits}
%
%
%
\bibliographystyle{splncs04}
\bibliography{refs}
\end{document}